\documentclass[a4paper]{article}
\usepackage{geometry}
\usepackage{amsmath,amsfonts,amsthm,amssymb}
\usepackage{wrapfig}
\usepackage{authblk}
\usepackage{fancyhdr}
\usepackage{chngpage}
\usepackage{color}
\usepackage{boxedminipage}
\usepackage{enumerate}
\usepackage{verbatim}    
\usepackage[square,sort,comma,numbers]{natbib}
\usepackage{hyperref}
\usepackage{graphicx, epstopdf} 
\usepackage{url}
\usepackage{appendix}
\usepackage[font=small,format=plain,labelfont=bf,up,textfont=it,up]{caption}
\usepackage{tikz}
\usetikzlibrary{arrows,automata}
\usepackage{listings}
\usepackage{upgreek}

\usepackage{subcaption}
\usepackage{appendix}
\usepackage{listings}
\usepackage{color}
\usepackage{lineno}
\definecolor{dkgreen}{rgb}{0,0.6,0}
\definecolor{gray}{rgb}{0.5,0.5,0.5}
\definecolor{mauve}{rgb}{0.58,0,0.82}

\title{Study of p-type silicon MOS capacitors at HL-LHC radiation levels through cobalt-60 gamma source and TCAD simulation}

\author{Patrick Asenov$^{*a \dagger}$, Panagiotis Assiouras$^{**a}$, Argiro Boziari$^{b}$, Konstantinos Filippou$^{c}$, Ioannis Kazas$^{a}$, Aristotelis Kyriakis$^{a}$, Dimitrios Loukas$^{a}$, Arianna Morozzi$^{d}$, Francesco Moscatelli$^{d, e}$ and Daniele Passeri$^{d, f}$\\
	{$^{*}$} E-mail: {patrick.asenov.asenov@cern.ch}\\ %, \email{second@mail.org},}
	
	{$^{**}$} E-mail: {panagiotis.assiouras@cern.ch}\\ %, \email{second@mail.org},}
	
	{$^{a}$} Institute of Nuclear and Particle Physics (INPP), NCSR Demokritos, Aghia Paraskevi, Greece\\
	{$^{b}$} Greek Atomic Energy Commission (GAEC),	Aghia Paraskevi, Greece\\
	
	{$^{c}$} University of Io\'annina, Io\'annina, Greece\\

	{$^{d}$} INFN Sezione di Perugia, Perugia, Italy\\

	{$^{e}$} Istituto Officina dei Materiali (IOM), Italian National Research Council (CNR), Perugia, Italy\\

	{$^{f}$} Department of Engineering, Universit\`a di Perugia, Perugia, Italy\\ 
	
	{$^{\dagger}$} Now at INFN Sezione di Perugia, Perugia, Italy and Istituto Officina dei Materiali (IOM), Italian National Research Council (CNR), Perugia, Italy.}

\begin{document}
	
	\maketitle
	\date

	\begin{abstract}
		During the era of the High Luminosity LHC (HL-LHC) the devices in its experiments will be subjected to increased radiation levels with high fluxes of neutrons and charged hadrons, especially in the inner detectors. A systematic program of radiation tests with neutrons and charged hadrons is being carried out by the CMS and ATLAS Collaborations in view of the upgrade of the experiments, in order to cope with the higher luminosity at HL-LHC and the associated increase in the pile-up events and radiation fluxes. In this work, results from a complementary radiation study with $^{60}$Co-$\upgamma$ photons are presented. The doses are equivalent to those that the outer layers of the silicon tracker systems of the two big LHC experiments will be subjected to. The devices in this study are float-zone oxygenated p-type MOS capacitors. The results of CV measurements on these devices  are presented as a function of the total absorbed radiation dose following a specific annealing protocol. The measurements are compared with the results of a TCAD simulation.
	\end{abstract}

	\section{Introduction}

		The High Luminosity Large Hadron Collider (HL-LHC) at CERN is expected to collide protons at a centre-of-mass energy of 14 TeV. It will reach the unprecedented peak instantaneous luminosity of $5-7.5 \mathrm{\times10^{34}\,cm^{-2}s^{-1}}$ with an average number of pileup events of 140-200. This will allow the LHC experiments to collect integrated luminosities up to 3000-4000$\, \mathrm{{fb}^{-1}}$ during the project lifetime~\cite{a} in order to search for new physics and study rare interactions. The increased statistics and extended physics reach however, come along with higher particle fluxes and total radiation doses which require more radiation-tolerant detectors and front-end electronics technologies. For this reason, new silicon tracking detectors with improved radiation hardness are needed for the HL-LHC experiments. 
		
		In the High Luminosity era the total absorbed doses in the outer layers of the tracking systems of the major LHC experiments are estimated to be in the order of 10-100$\,$kGy, depending on the distance from the beam line. A systematic campaign of irradiation tests with neutrons and charged hadrons initiated by the LHC collaborations is ongoing in order to estimate how will the tracking systems cope with the demands for higher luminosity and radiation fluxes. Complementary radiation studies with $^{60}$Co-$\upgamma$ photons are performed with doses equivalent to those that the outer layers of the silicon tracker systems of the two large LHC experiments will be subjected to, along with the related TCAD simulations. One of these studies is presented here.
	
	\section{Samples and laboratory equipment}

		The samples used for irradiation in this study are float-zone oxygenated silicon n-in-p test structures from thinned 240$\, \upmu$m thick wafers produced by Hamamatsu Photonics K.K.~\cite{b}. Each test structure contains one square MOS (area = 4$\,$mm $\times$ 4$\,$mm). 

\begin{figure}[!htbp]
			\centering % \begin{center}/\end{center} takes some additional vertical space
			\includegraphics[width=0.45\textwidth,height=6.0cm,origin=c,angle=0]{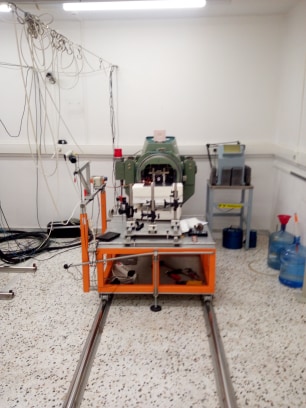}
			\qquad
			\includegraphics[width=0.45\textwidth,height=6.0cm,origin=c,angle=0]{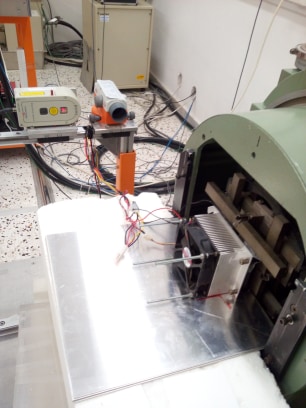}
			% "\includegraphics" from the "graphicx" permits to crop (trim+clip)
			% and rotate (angle) and image (and much more)
			\caption{\label{figSetup} Left: The Cobalt-60 source: Picker therapy unit. Right: The container with the samples in front of the source. The fan and the	thermoelectric cooler are visible.}
		\end{figure}

		Cobalt-60 has two characteristic gamma-ray decay modes with energies 1.1732 MeV and 1.3325$\,$MeV, respectively. These energies are much harder than those used in common X-ray irradiation tests. The $^{60}$Co source used in the current study is a  Picker teletherapy unit~\cite{picker} with a radioactivity of 30$\,$TBq as of March 2012, estimated at approximately 11$\,$TBq by the time the measurements were performed, with a horizontal orientation (Figure~\ref{figSetup}, left). It was calculated by using FC65-P Ionization Chambers from IBA Dosimetry~\cite{c} that the dose rate at irradiation point (40$\,$cm from the source) is 0.96$\,$kGy/h. The irradiation was performed in the secondary standard ionizing radiation laboratory of the Greek Atomic Energy Commission (GAEC), accredited according to ISO 17025 in the field of radiotherapy, and the relevant CMCs (calibration and measurement capabilities) are published in the BIPM database~\cite{d}.
		
\begin{figure}[!htbp]
			\centering % \begin{center}/\end{center} takes some additional vertical space
			\includegraphics[width=0.4\textwidth,height=6.0cm,origin=c,angle=0]{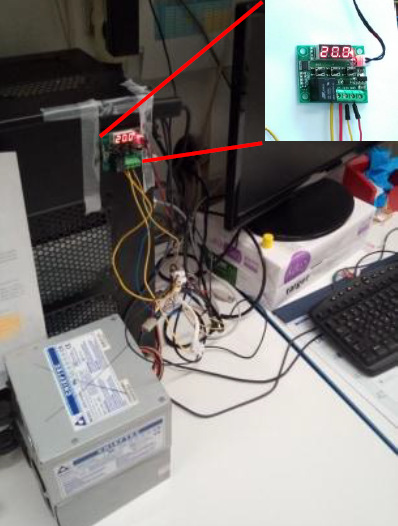}
			\qquad
			\includegraphics[width=0.4\textwidth,height=6.0cm,origin=c,angle=0]{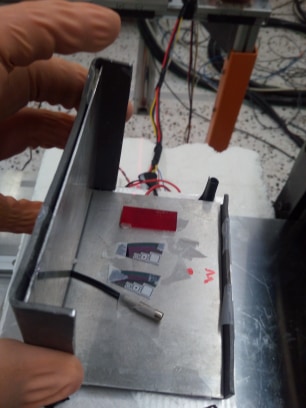}
			% "\includegraphics" from the "graphicx" permits to crop (trim+clip)
			% and rotate (angle) and image (and much more)
			\caption{\label{figCPE} Left: The microcontroller and power supplies of the irradiation setup. Right: The lead-aluminum container for charged particle equilibrium (CPE).}
		\end{figure}

\begin{figure}[!htbp]
			\centering % \begin{center}/\end{center} takes some additional vertical space
			\includegraphics[height=7.0cm,width=0.7\textwidth,origin=c,angle=0]{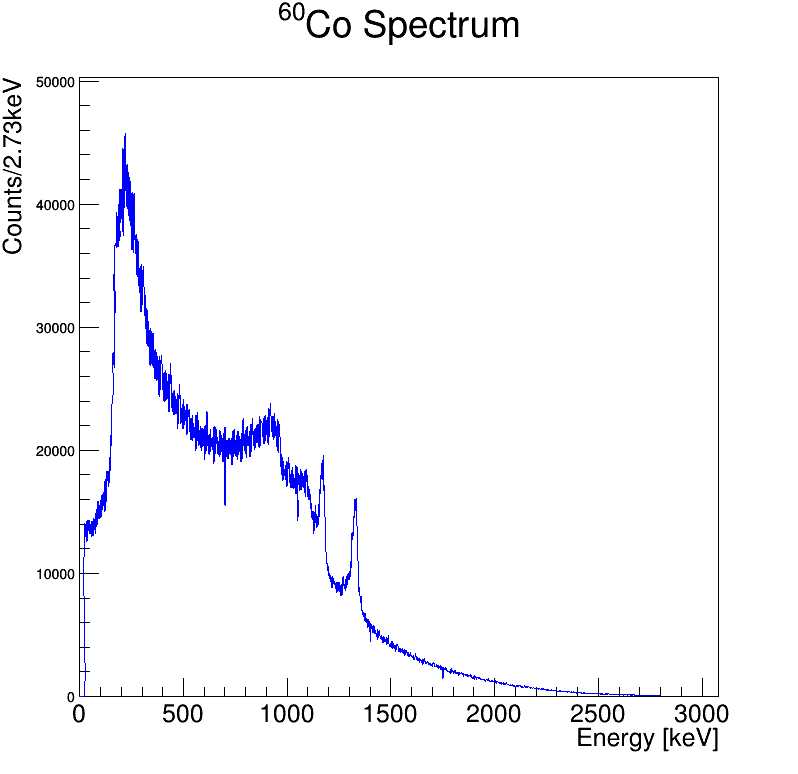}
			\caption{\label{figSpec} Energy spectrum taken 4$\,$m away from the cobalt-60 source and after 5.2$\,$cm of Pb (a 5$\,$cm thick Pb block was placed between the source and the CPE box and the Pb thickness of the CPE box was 0.2$\,$cm), as measured inside the charged particle equilibrium (CPE) box. The left peak is a backscatter peak at approximately 200$\,$keV which emerges when $\upgamma$-rays enter the material around the detector and are scattered back into the detector. The right peaks are the peaks corresponding to the 1.1732$\,$MeV and 1.3325$\,$MeV characteristic gamma-ray decay modes of cobalt-60.}
		\end{figure}

		The cooling system consisted of a thermoelectric cooler (Peltier element, type TEC1 12704) operating at temperature lower than room temperature (8$\, ^\circ$C $\pm$ 1$\, ^\circ$C), an aluminum plate and a fan for heat dissipation (Figure~\ref{figSetup}, right). The selected Peltier is sealed with 704 silicon rubbers and proved to be robust against $\upgamma$-irradiation from the Cobalt-60 source. A microcontroller for the stabilization of temperature and the respective power supplies were used in addition (Figure~\ref{figCPE}, left). Charged particle equilibrium (CPE) was achieved due to a box of 2$\,$mm thick Pb and 0.8$\,$mm of inner lining Al sheet, i.e. a lead-aluminum container for the absorption of low energy photons and secondary electrons~\cite{e} where the samples were kept during irradiation (Figure~\ref{figCPE}, right). The energy spectrum inside the CPE container, 4$\,$m away from the source, was measured (Figure~\ref{figSpec}) with a Micro-sized CZT Gamma Spectrometer with a volume of 0.5$\,$cm$^3$, a spectral response in the range 30$\,$keV - 3$\,$MeV and an energy resolution $<\: 2.5$\% at 662$\,$keV ($^{137}$Cs)~\cite{mgs}. As for the calculation of absorbed doses in silicon and $SiO_{2}$, it should be noted that this is quite straightforward since for $\upgamma$-rays of energies ranging from 200$\,$keV to 2$\,$MeV (where the bulk spectrum of the used $^{60}$Co source lies within) conversion from gray $[Gy]$ in Air to gray $[Gy]$ in Silicon is simply done using a unity  multiplication factor ~\cite{nist}, ~\cite{us_doh}.

		Electrical measurements were performed in the range 15$\, ^\circ$C - 20$\, ^\circ$C (and below) using an automatic probe station (Carl Suss PA 150) and supplementary equipment (HP4192A, Keithley 6517A) for electrical characterization of microelectronic devices and the samples were annealed in a Weiss WKS 3-180/40/5 climate test chamber. Data analysis was subsequently performed using the ROOT analysis software~\cite{root}.
		
	\section{Experimental procedure and protocol}
		
		The CPE container with the samples was held 40$\,$cm away from the source while being irradiated. The irradiation was split in slots of approximately 14-16 hours of irradiation. After each slot, annealing of the samples was performed in the climate test chamber at 60$\,^\circ$C for 10 min (corresponding to four days of annealing at room temperature). The electrical tests at the probe station after the annealing were performed at 15-23$\,^\circ$C using LabVIEW as data taking and control software. Humidity in the lab was controlled with a desiccator and during all measurements RH was below 30\%. The oscillation amplitude for the CV measurements was set at 250$\,$mV. CV measurements were carried out for various frequencies (100$\,$Hz, 1$\,$kHz, 10$\,$kHz, 100$\,$kHz, 1$\,$MHz for MOS capacitors).
		
	\section{Experimental results of MOS capacitors (CV analysis)}	
	
\begin{figure}[!htbp]
			\centering
			\includegraphics[width=0.6\textwidth,height=9.0cm]{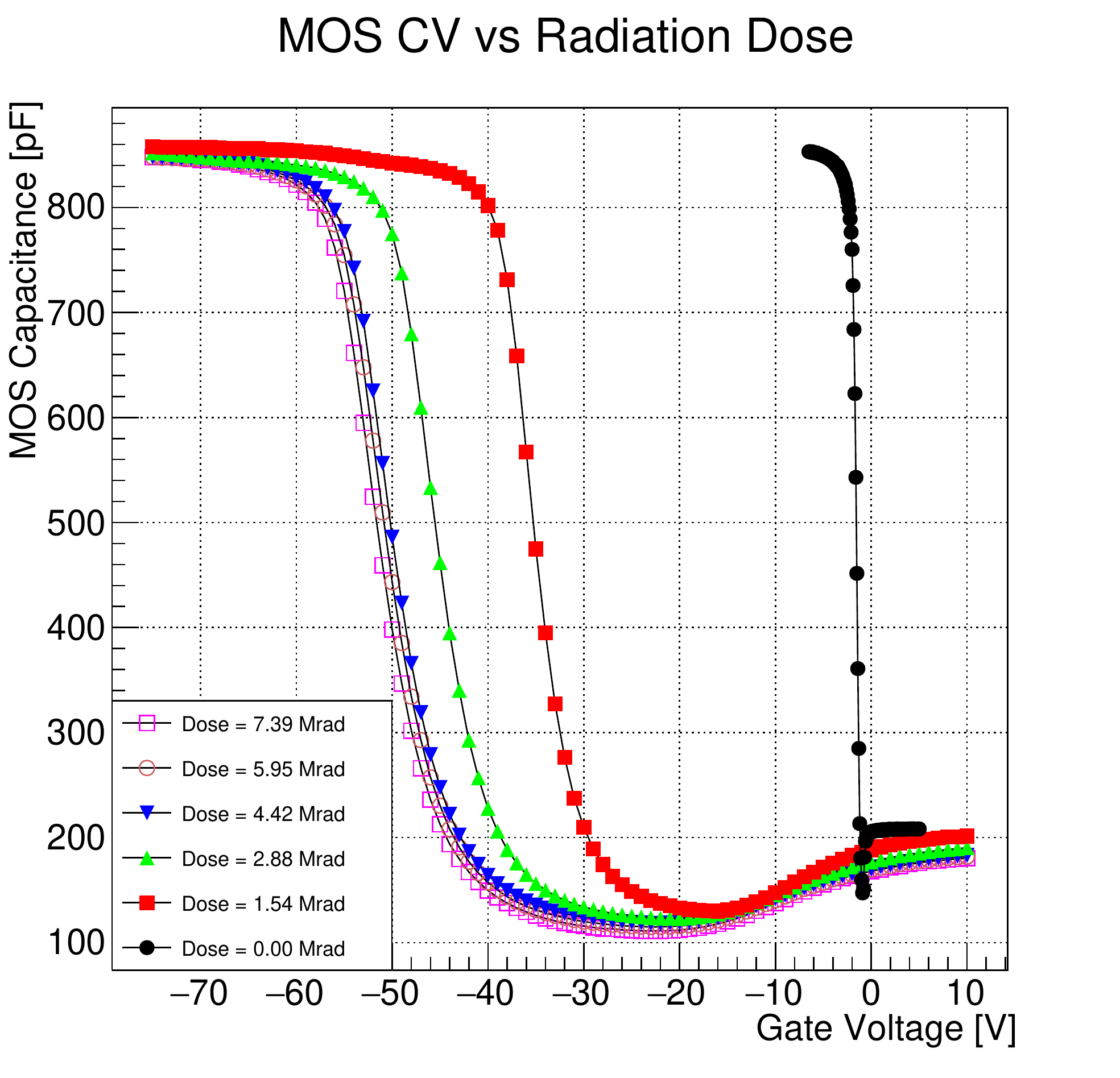}
			\caption{Capacitance-voltage curves for a MOS capacitor for various doses; measurement frequency = 10$\,$kHz. Irradiation time slots of 14-16 hours, no removal of the samples from the cobalt-60 source between the irradiation time slots, control over humidity during electrical measurements.}
			\label{figP2MOSCVDoses}
		\end{figure}

\begin{figure}[!htbp]
			\centering % \begin{center}/\end{center} takes some additional vertical space
			\includegraphics[width=0.8\textwidth,height=9.0cm,origin=c,angle=0]{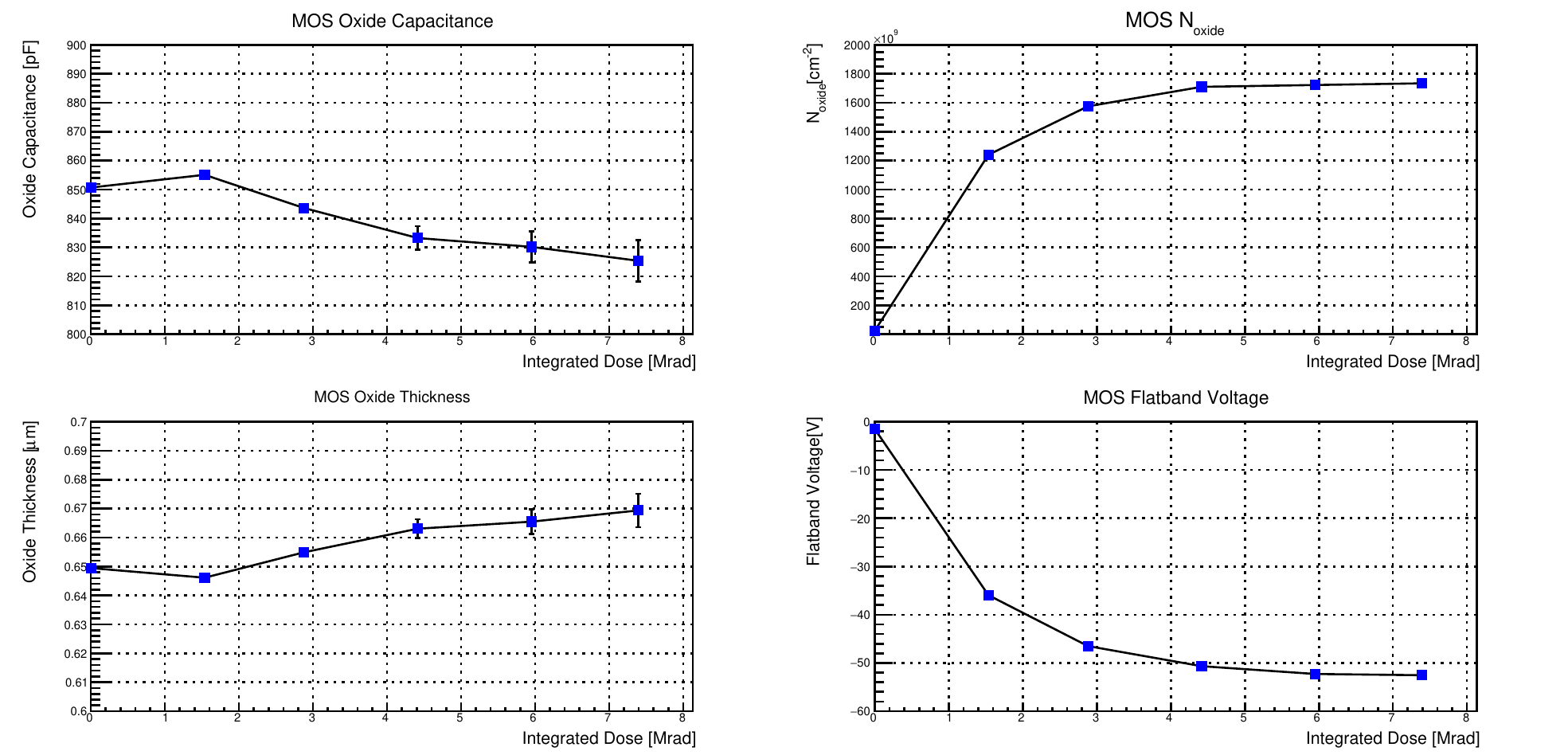}
			\caption{Various features of a MOS capacitor before and after irradiation; measurement frequency = 10$\,$kHz. Top left: Oxide capacitance.  Top right: Effective oxide concentration. Bottom left: Oxide thickness. Bottom right: Flatband voltage. Irradiation time slots of 14-16 hours, no removal of the samples from the cobalt-60 source between the irradiation time slots, control over humidity during electrical measurements.}
			\label{figP2MOSchar1} 
		\end{figure}

		Figure~\ref{figP2MOSCVDoses} shows a typical capacitance-voltage curve of a MOS capacitor for various doses. After the exposure to gamma photons there is a clear evidence of positive charge induced in the oxide of the p-type MOS capacitors. Various interesting parameters can be extracted from this figure, such as the oxide capacitance ($C_{oxide}$), the effective oxide concentration ($N_{oxide}$), the oxide thickness ($t_{oxide}$) and the flatband voltage ($V_{FB}$), as can be seen in Figure~\ref{figP2MOSchar1}. The flatband voltage is calculated by using the maximum of the first derivative of the MOS CV curve, which yields the flatband voltage as the inflection point of the curve. To improve the performance of this method, a cubic spline interpolation is performed on the data before calculating the first derivative. The oxide capacitance is considered to be the capacitance measured in accumulation, which is the intersection point of a linear fit with slope fixed to zero to the data points in the accumulation region with the capacitance axis. The oxide thickness is calculated from the oxide capacitance ($C_{oxide}$) using the following relation:
		\begin{equation}
			\label{eq:1}
			t_{oxide} = \frac{\epsilon_{oxide} A_{gate}}{C_{oxide}}
		\end{equation}
		\linebreak where $\epsilon_{oxide} = 0.3453\ pF/cm$ is the permittivity of the oxide material and $A_{gate} = 0.16\,$cm$^2$ is the gate area of the MOS device.
		The oxide thickness is almost flat within errors.
		
		 Finally, the effective oxide concentration is calculated as follows:
		\begin{equation}
			\label{eq:2}
			N_{oxide} = \frac{C_{oxide}}{q A_{gate}} (\phi_{ms} - V_{FB})
		\end{equation}
		\linebreak where $\phi_{ms}$ is the work function difference between the aluminum gate layer and p-type silicon  varying slightly as a function of the doping concentration as a consequence of the irradiation of our sample.
		
		It is observed that there is a saturation of the value of the flatband voltage at around $-52\,$V. This leads to stabilization of $\mathrm{N_{oxide}}$ after the initial increase even at high doses.

	\section{TCAD simulation of MOS capacitors}
		
		Technology computer aided design (TCAD) \cite{TCAD} is used in order to provide a better insight of the complex phenomena related to semiconductor devices. It follows a numerical modeling approach and it can be used in order to simulate the fabrication procedure of new devices, operation and reliability under real conditions or even endurance of the devices in harsh environments. As the next generation of silicon sensors for the HL-LHC era will have to withstand unprecedentedly high fluences, TCAD simulations and experimental studies have been performed in order to provide a better understanding of the mechanisms of radiation damage in silicon devices \cite{Moll}, \cite{Schwandt}, \cite{Passeri1}, \cite{Petasecca}, \cite{Dalal}.

		\subsection{Description of the radiation model}
		
		Particles passing through the silicon bulk produce radiation effects in the medium, some of which are reversible and other irreversible. In principle, the radiation damage in silicon sensors is caused by two different factors: the ionizing energy loss and the non-ionizing energy loss. Although the ionizing energy loss is important for the signal formation and is usually reversible, it can cause irreversible effects on the oxide by introducing positive oxide charge in the SiO$_{2}$, by increasing the number of bulk oxide traps and by increasing the number of interface traps. These effects are usually referred to \emph{surface damage}. This leads to conducting layers in silicon which influence the operation of segmented silicon sensors with respect to the inter-electrode isolation, the breakdown voltage and the charge collection efficiency. The non-ionizing energy loss is responsible for the introduction of defects into the silicon lattice through the displacement of crystal atoms, usually produced due to the impact of high-momentum particles. These impacts lead to point and cluster defect generation and hence to the introduction of deep-level trap states which act like generation-recombination centers \cite{Moll}. Non-ionizing effects are referred to as \emph{bulk damage} and on a macroscopic scale they are responsible for the increase of the leakage current in silicon sensors, the changes in the effective space charge concentration and the charge collection efficiency.	
		
		In \cite{Passeri1} and \cite{Petasecca} a three level model is presented for simulating the bulk damage effects for n-type and p-type substrates, which is usually referred to as the \emph{Perugia model}. In order to incorporate also the surface damage effects an extension of the model has been made refereed to as the \emph{Perugia surface model 2019} presented in \cite{mosc}, \cite{mor}. The surface damage effects can be mainly described by two parameters: the oxide charge ($Q_{ox}$) and the interface trap states ($N_{IT}$ ). The values of the above quantities can be extracted from high-frequency and quasi-static CV measurements by following the procedure described in \cite{Nicollian}.  The model is able to reproduce the radiation damage macroscopic effects up to the order of $10^{15}\, \mathrm{n_{eq}/cm^{2}}$ 1 MeV neutron equivalent fluences and X-ray photon doses of up to $10$ $Mrad$. 
		
		The  three variables, oxide charge density $Q_{ox}$ [$\mathrm{cm^{-2}}$], acceptor integrated interface trap state density $N_{IT_{acc}}$ [$\mathrm{cm^{-2}}$] and donor integrated interface trap state density $N_{IT_{don}}$ [$\mathrm{cm^{-2}}$] are used as input parameters for the simulation where the macroscopic factors, such as the capacitance or the leakage current, are calculated. They are related to the surface damage effects, vary with the dose/fluence and can be represented by the following equations, according to \cite{mor}:
		
		\begin{align}
		Q_{ox}( x ) &= Q_{ox}(0)+\Delta Q_{ox}( x ) \\ 
		N_{IT_{acc}}( x ) &= N_{IT_{acc}}(0)+\Delta N_{IT_{acc}}( x ) \\
		N_{IT_{don}}( x ) &= N_{IT_{don}}(0)+\Delta N_{IT_{don}}( x )
		\end{align}
		
		where $Q_{ox}(0)$ , $N_{IT_{acc}}(0)$ and $N_{IT_{don}}(0)$ are the values before irradiation and $\Delta Q_{ox}( x )$, $\Delta N_{IT_{acc}}( x )$ and $\Delta N_{IT_{acc}}( x )$ are the values after irradiation, where $x$ is the dose in Mrad.

	    The radiation model that is used in this work is a modified version of the surface damage model presented in \cite{mor}. It also relies on two uniform defect energy band distributions: one which accounts for the acceptor-like interface trap states near the conduction band and another one which accounts for the donor-like interface defects near the valence band. Table \ref{tab:surface_damage_parameters} summarizes the energy band range ($E_{T}$) for the acceptor and donor states, the band width and the trap capture cross section of electron  $\sigma_{e}$ and holes $\sigma_{h}$ that are used in our modified  model, in accordance to \cite{mor}. Due to the fact that the MOS capacitance characteristics are strongly affected by the complex phenomena taking place in the SiO$_{2}$ region and the Si-SiO$_{2}$ interface, only the surface damage effects are taken into consideration.

			\begin{table}[!htbp]
			\caption{The energy band range ($E_{T}$) for the acceptor and donor states, the band width and the trap capture cross section of electron $\sigma_{e}$ and holes $\sigma_{h}$ that are used in our modified \emph{Perugia surface model 2019}, in accordance to \cite{mor}. $E_{V}$ and $E_{C}$ refer to the Valence and Conduction energy levels respectively. }			
			\begin{footnotesize}
				\label{tab:surface_damage_parameters}
				\begin{center}
				\begin{tabular}{||c c c c c||} 
					\hline
					Type & Energy (eV) & Band width (eV) & $\sigma_{e}$ ($cm^{2}$) & $\sigma_{h}$ ($cm^{2}$) \\ [0.5ex] 
					\hline\hline
					Donor & $E_{V}<E_{T}<E_{V}+0.54$ & 0.54 & $1.0 \times 10^{-15}$  & $1.0 \times 10^{-16}$ \\ 
					\hline
					Acceptor & $E_{C}-0.58<E_{T}<E_{c}$ & 0.58 &$1.0 \times 10^{-16}$  & $1.0 \times 10^{-15}$ \ \\
					\hline
				\end{tabular}
				\end{center}
			\end{footnotesize}
		   \end{table}

		%For simulating the bulk damage, three deep-level traps are introduced in addition to the already mentioned traps for the surface damage simulation \cite{Passeri1}, %\cite{Passeri2}. The energy, the introduction rate ($\eta$), which is the factor that multiplies the flux giving the concentration of the deep level traps and the cross %section of electron  ($\sigma_{e}$) and holes ($\sigma_{h}$) of the three trap levels are reported in Table \ref{tab:bulk_damage_parameters}, according to \cite{mor}. 
		%
	%	
	%	\begin{center}
	%		\begin{table}
	%		\caption{The energy, the introduction rate ($\eta$) and the cross section of electron ($\sigma_{e}$) and holes ($\sigma_{h}$) of the three trap levels are reported in %Table \ref{tab:bulk_damage_parameters}, according to \cite{mor}. $E_{C}$ refer to the Conduction energy levels. }
%			
%			%\begin{footnotesize}
%				\label{tab:bulk_damage_parameters}
%				\begin{tabular}{||c c c c c||} 
%					\hline
%					Type &  Energy (eV) &$\eta (cm^{-1})$ & $\sigma_{e}$ ($cm^{2}$) & $\sigma_{h}$ ($cm^{2}$) \\ [0.5ex] 
%					\hline\hline
%					Donor & $E_{C}-0.23$ & $0.06$ & $2.3 \times 10^{-14}$  & $2.3 \times 10^{-15}$ \ \\
%					\hline
%					Acceptor & $E_{C}-0.42$  & $1.613$  & $1.0 \times 10^{-15}$ & $1.0 \times 10^{-14}$ \\ 
%					\hline
%					Acceptor & $E_{C}-0.46$  & $0.09$  & $7.0 \times 10^{-15}$ & $7.0 \times 10^{-14}$ \\ 
%					\hline
%				\end{tabular}
%			%\end{footnotesize}
%		    \end{table}
%		\end{center}

		\subsection{Description of the TCAD simulations}
		
\begin{figure}[!htbp]
			\centering
			\begin{subfigure}{0.48\textwidth}
				\centering
				\includegraphics[height=4.5cm,width=\textwidth]
				{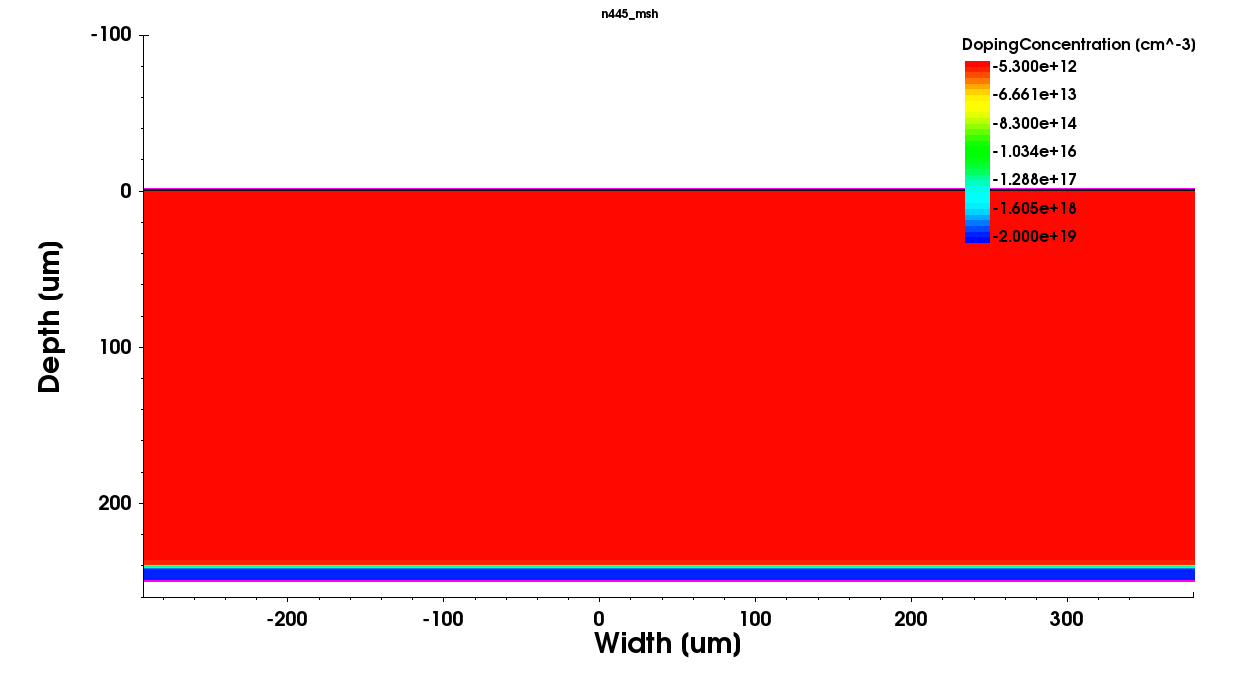}
				\caption{\label{fig:MOS_Sim_(-500_to_500)}}      
			\end{subfigure}
			\quad
			\begin{subfigure}{0.48\textwidth}
				\centering
				\includegraphics[height=4.5cm , width=\textwidth]{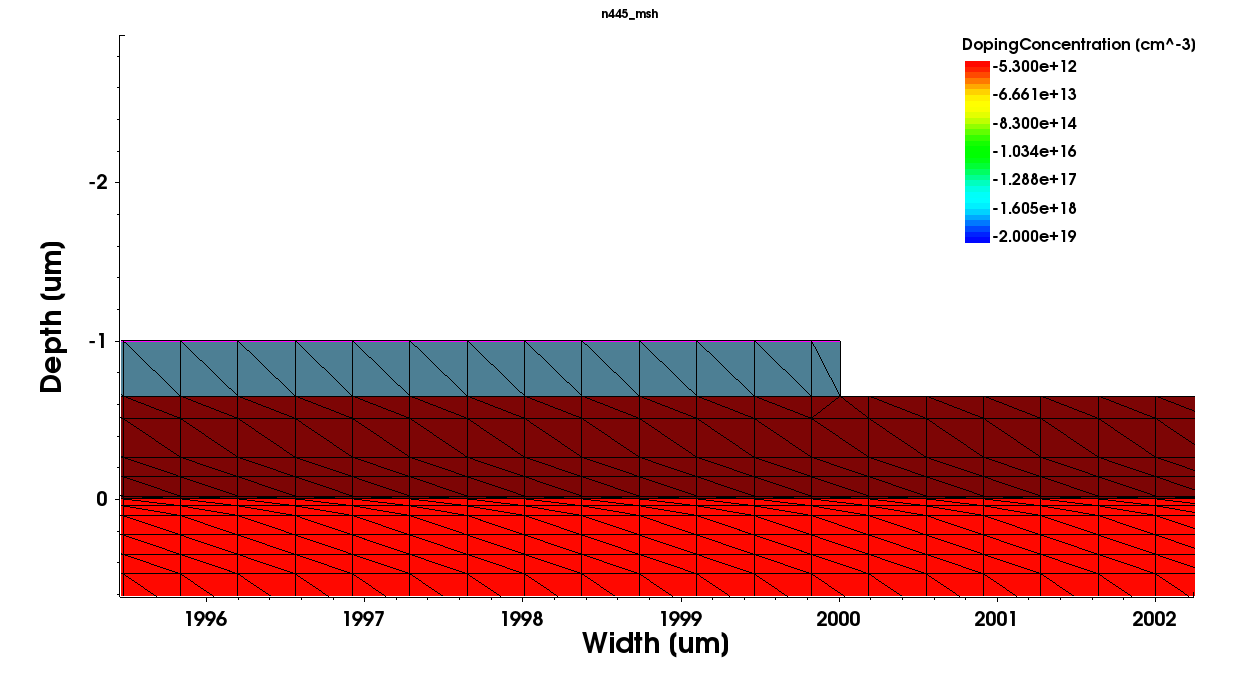}
				\caption{ \label{fig:MOS_Sim_close_up}}
			\end{subfigure}	
			\label{fig:Abs.Rel.Diff}
			\caption[]{The structure that is used for the 2D simulation of the MOS capacitors: a segment of the structure depicting the depth of the whole structure (Figure \ref{fig:MOS_Sim_(-500_to_500)}) and a close-up view of the simulated structure near a metal edge (Figure \ref{fig:MOS_Sim_close_up}). The color variation depicts the doping concentration. The p-type implant is displayed in red, the aluminum contacts are displayed in gray and the SiO$_{2}$ is displayed in dark red.}	
		\end{figure}

		Figures ~\ref{fig:MOS_Sim_(-500_to_500)} and \ref{fig:MOS_Sim_close_up} show a section of the 2D structure that is used in this work for the simulation of the MOS capacitors. The silicon substrate is indicated in light red, the oxide is indicated in dark red, and the aluminum metal is indicated in gray. The geometrical characteristics that were used for the simulation are summarized in the Appendix \ref{App:TCAD_geometrical properties}.

		In order to calculate the capacitances, a small signal AC analysis is performed at 10$\,$kHz. This frequency corresponds to the frequency at which the experimental measurements have been performed. Some of the physical models that are used in this work are the Auger recombination, Shockley-Read-Hall recombination, avalanche electron-hole generation, doping dependence mobility and high field saturation \cite{Sentaurus}. The temperature in the simulations was set at 293$\,$K which corresponds to the experimental temperature during measurements. The physical models used for the simulations are shown in Appendix \ref{App:TCAD_physical_models}

          \begin{equation}
             \label{eq:6}
             \Delta Q_{ox}( x ) =  1.08\cdot10^{12} + 3.41\cdot10^{11} ln( x ) \, \quad [\mathrm{cm^{-2}}]
           \end{equation} 
           \begin{equation}
             \label{eq:7}
             \Delta N_{IT_{acc}}( x ) = 2.30\cdot10^{12} + 2.65\cdot10^{11} ln( x ) \, \quad [\mathrm{cm^{-2}}]
           \end{equation} 
           \begin{equation}
             \label{eq:8}
             \Delta N_{IT_{don}}( x ) = 4.26\cdot10^{11} + 1.98\cdot10^{11} ln( x ) \, \quad [\mathrm{cm^{-2}}]
           \end{equation}

		The oxide charge concentration is shown in Figure \ref{figP2MOSchar1} as it has been calculated using equation \ref{eq:2}. These values have been set as inputs for the $Q_{ox}$ parameter. The non-irradiated term  $Q_{ox}(0)$ is set equal to $2.4\cdot10^{10} \, \mathrm{cm^{-2}}$ and the irradiated one $\Delta Q_{ox}( x )$ is set according to equation \ref{eq:6} ($x$ is the total dose in Mrad) and are extracted from measurements as can be seen in Figure \ref{figP2MOSchar1}.
		
		 The parameters $N_{IT_{acc}}(0)$ and $N_{IT_{don}}(0)$ are set equal to $2.0\cdot10^{9} \, \mathrm{cm^{-2}}$ according to \cite{mor1}.
		 
\begin{figure}[!htbp]
			\centering
			\makebox[\textwidth][c]{\includegraphics[height=5.5cm, width=0.6 \textwidth,origin=c,angle=0]{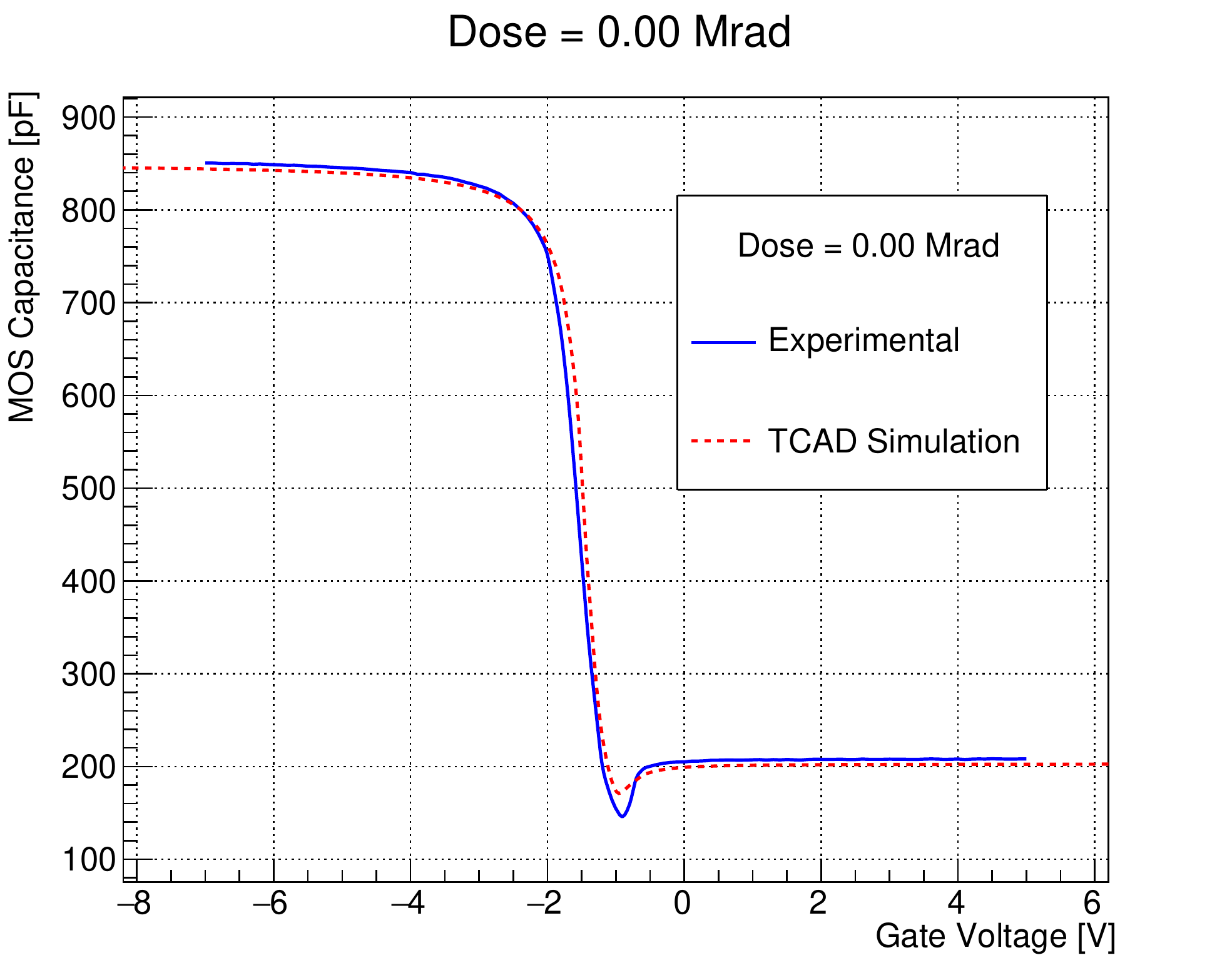}}
			\caption{TCAD Simulated CV of non-irradiated MOS capacitor compared to the experimental measurements.}
			\label{fig:SimulationUnirradiated}
		\end{figure}

		 Figure \ref{fig:SimulationUnirradiated} shows the simulated CV results for the non-irradiated MOS capacitor in comparison with the experimental data. The matching is extremely good as can be seen.
		
\begin{figure}[!htbp]
			\centering
			\begin{subfigure}{0.48\textwidth}
				\centering
				\includegraphics[height=6.0cm,width=\textwidth]
				{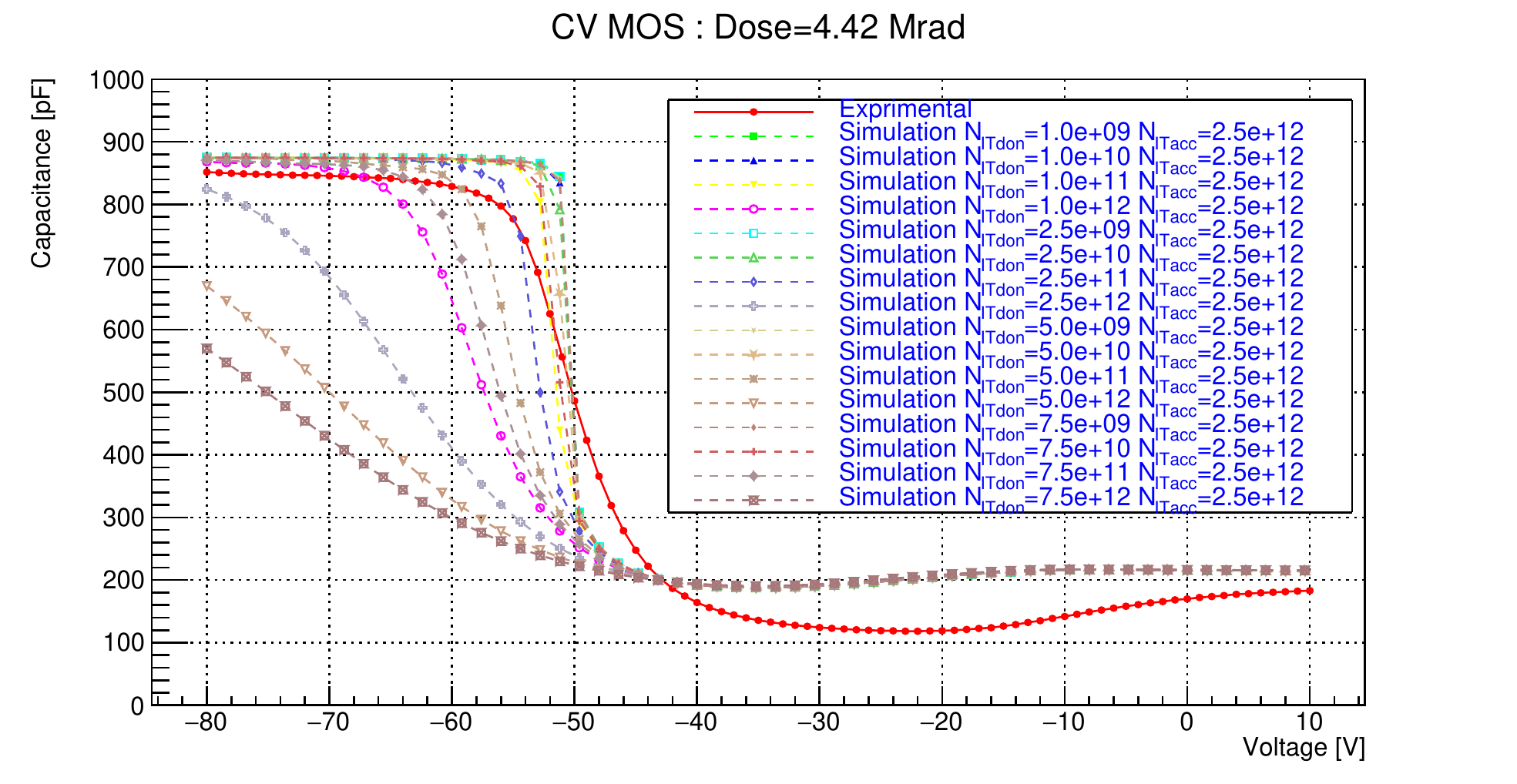}
				\caption{\label{fig:SimulationIrradiated_Nitacc_Stable}}      
			\end{subfigure}
			\quad
			\begin{subfigure}{0.48\textwidth}
				\centering
				\includegraphics[height=6.0cm  , width=\textwidth]{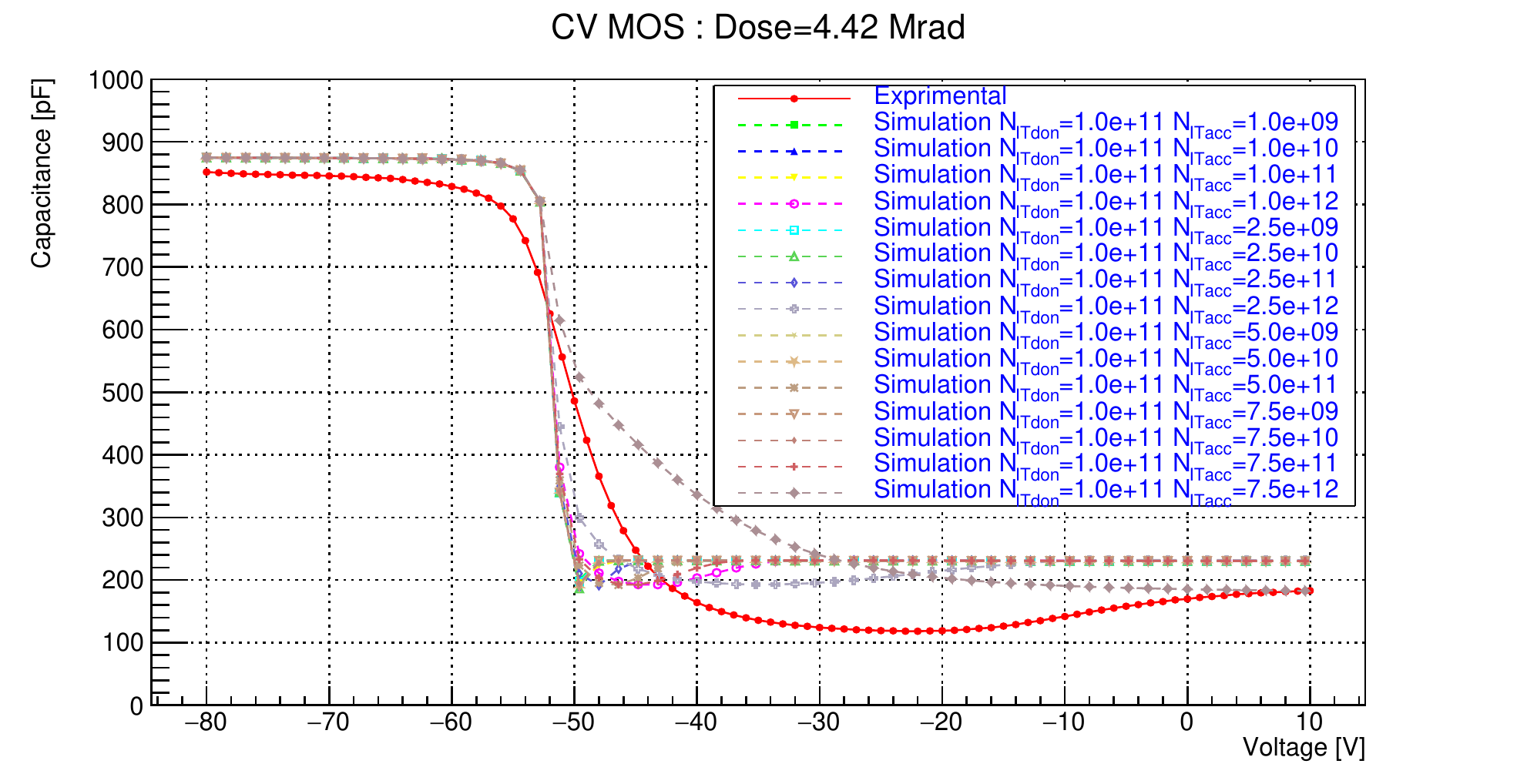}
				\caption{ \label{fig:SimulationIrradiated_Nitdon_Stable}}
			\end{subfigure}	
			\label{fig:fig:Abs.Rel.Diff}
			\caption[]{Simulated CV plots at Dose = 4.42$\,$ Mrad by keeping $N_{IT_{acc}}$ stable and changing the $N_{IT_{don}}$ (Figure \ref{fig:SimulationIrradiated_Nitacc_Stable}) and by keeping $N_{IT_{don}}$ stable and changing $N_{IT_{acc}}$ (Figure \ref{fig:SimulationIrradiated_Nitdon_Stable}), respectively, compared to  the experimental results at the same dose level.}	
		\end{figure}

\begin{figure}[!htbp]
		   \centering
	    	\makebox[\textwidth][c]{\includegraphics[height=4.0cm, width=1.\textwidth,origin=c,angle=0]{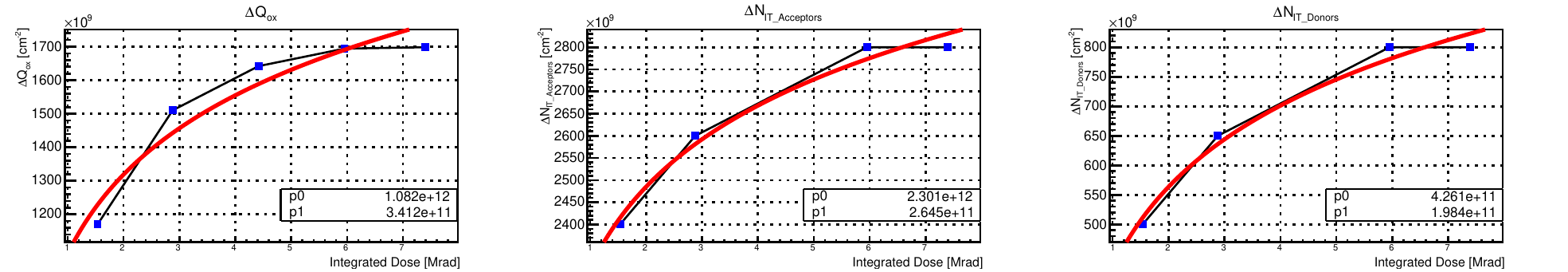}}
		   \caption{The $\Delta Q_{ox}( x )$(left), $\Delta N_{IT_{acc}}( x )$(middle) and $\Delta N_{IT_{don}}( x )$(right) values that best match our experimental data. Both are expressed in the form of p0 + p1$\,$ln(x), where p0 and p1 are in cm$^{-2}$ and \emph{x} is the dose in \emph{Mrad}.}
	    	\label{fig:DeltaNit_Model}
	     \end{figure}

		Although experimental measurements haven't been performed for extracting the values of $N_{IT_{acc}}$ and $N_{IT_{don}}$ in this work, several cases of these values have been investigated. Figure \ref{fig:SimulationIrradiated_Nitacc_Stable} shows the TCAD simulated CV plots at an indicative total dose of 4.42$\,$ Mrad by keeping $N_{IT_{acc}}$ stable at $2.5\cdot10^{12} \, \mathrm{cm^{-2}}$ and by changing the $N_{IT_{don}}$ from $1.0\cdot10^{9} \, \mathrm{cm^{-2}}$ to $7.5\cdot10^{12} \, \mathrm{cm^{-2}}$, compared with the experimental results at the same dose level. In a similar manner, Figure \ref{fig:SimulationIrradiated_Nitdon_Stable} shows the TCAD simulated CV plots at the same dose by keeping $N_{IT_{don}}$ stable at $1.0\cdot10^{11} \, \mathrm{cm^{-2}}$ and by changing $N_{IT_{acc}}$ from $1.0\cdot10^{9} \, \mathrm{cm^{-2}}$ to $7.5\cdot10^{12} \, \mathrm{cm^{-2}}$. Figure  \ref{fig:DeltaNit_Model} shows the $\Delta Q_{ox}( x )$(left), $\Delta N_{IT_{acc}}( x )$(middle) and $\Delta N_{IT_{don}}( x )$(right) values that better match our experimental data. The fitted parameters are shown in equations \ref{eq:6}, \ref{eq:7} and \ref{eq:8} for $\Delta Q_{ox}( x )$, $\Delta N_{IT_{acc}}( x )$ and $\Delta N_{IT_{don}}( x )$ ($x$ is the dose in Mrad) respectively.
		
\begin{figure}[!htbp]
			\centering
			\makebox[\textwidth][c]{\includegraphics[height=16cm, width=0.9\textwidth,origin=c,angle=0]{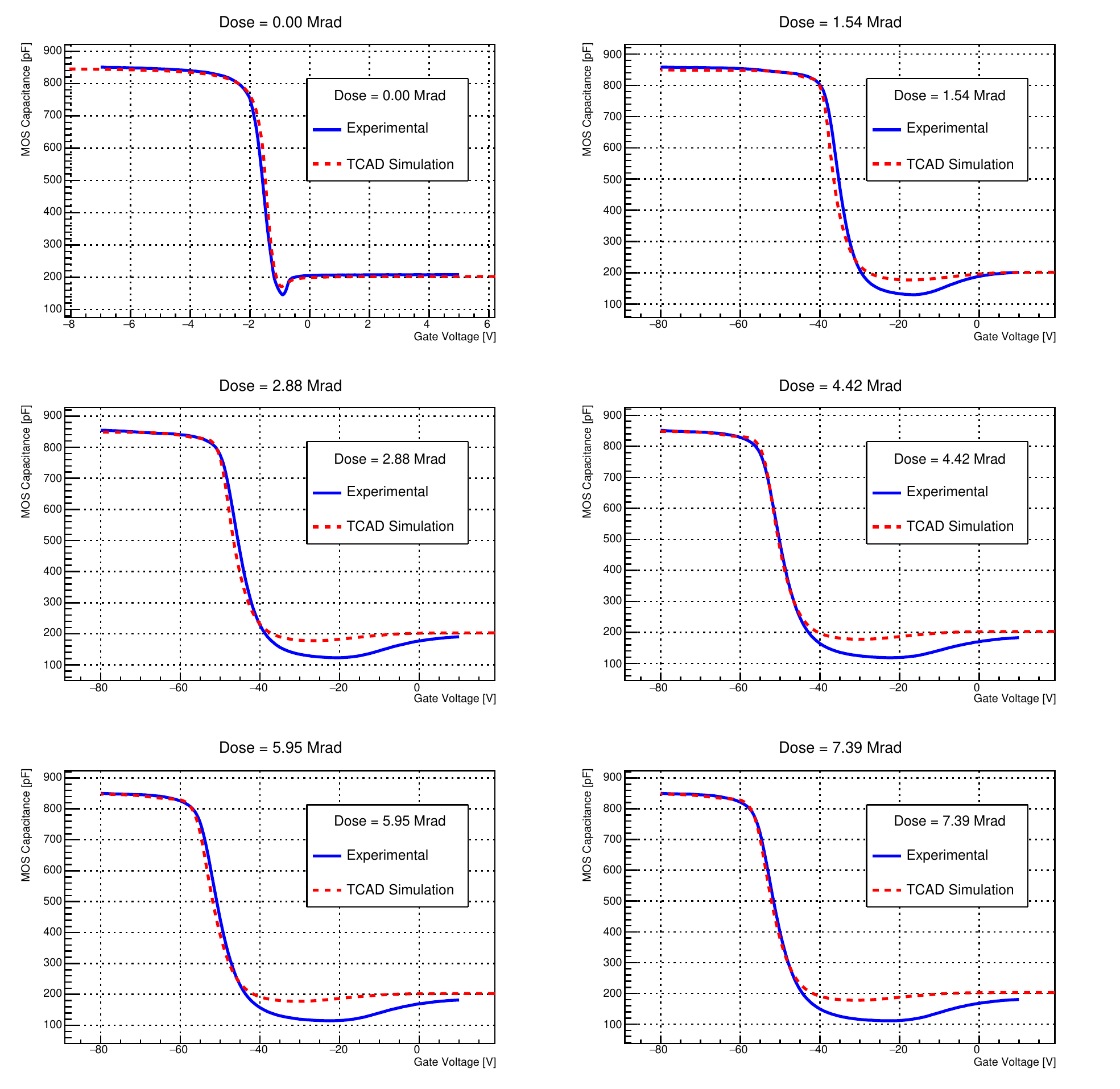}}
			\caption{Experimental and TCAD simulated CV curves for a MOS capacitor for various doses; measurement frequency = 10$\,$kHz. The model describes well the experimental data.}
			\label{figSimulation}
		\end{figure}

	      Figure \ref{figSimulation} shows the experimental (solid line) and simulated (dashed line) MOS CV characteristics for various doses. The  $\Delta N_{IT_{acc}}( x )$ and $\Delta N_{IT_{don}}( x )$ values for a total dose of $ 4.42 \,$Mrad was evaluated from our model given by equations \ref{eq:6}, \ref{eq:7} and \ref{eq:8}. As can be seen, the TCAD simulation based on our modified \emph{Perugia surface model 2019} describes well our $^{60}Co$ irradiation measurements. 		
		
	\section{Conclusions}

		In this work silicon MOS capacitors  were irradiated with $^{60}$Co-$\upgamma$ photons from a $\sim$11 TBq source. The total absorbed dose was $\sim$7.4 Mrad. The level of the radiation-induced charge in the test structures was determined from the shift of the flatband voltage in the MOS capacitors after irradiation and a saturation effect was observed. Apart from the flatband voltage, the irradiation of the MOS capacitors showed significant change in the threshold voltage and depletion region slope, which is related to the charge concentration. The measurements were compared with the results of a TCAD simulation based on a modified version of the  \emph{Perugia  model 2019}, which takes into account several radiation damage effects. The model describes well our experimental measurements.
		
	\newpage
	\appendix
	\section{Geometrical properties used for the TCAD simulation}
	\label{App:TCAD_geometrical properties}
	The following geometrical properties were used for the simulation of the 2D MOS structure with the TCAD Synopsys:
	
	\begin{table}[htp!]
		\begin{footnotesize}
			\centering
			\begin{tabular}{|c|c|} 
				\hline
				\begin{tabular}[c]{@{}l@{}} Detector thickness \end{tabular} & 250 $\upmu$m \\
				\hline
				\begin{tabular}[c]{@{}l@{}} Backplane thickness \end{tabular} & 10 $\upmu$m \\
				\hline
				\begin{tabular}[c]{@{}l@{}} $N_{bulk}$ \end{tabular} & $5.3 \times 10^{12}$ cm$^{-3}$ \\
				\hline
				\begin{tabular}[c]{@{}l@{}} Oxide thickness \end{tabular} & 0.65 $\upmu$m \\
				\hline
			\end{tabular}
			\caption{Geometrical properties and doping concentration used for the TCAD simulation of the MOS capacitors. }
			\label{table1}
		\end{footnotesize}
	\end{table}
	
	\section{Physical models used for the TCAD simulation}
	\label{App:TCAD_physical_models}
	
	The following physical models were used for the simulations in Synopsys TCAD:

	\begin{lstlisting}
		Physics{
			AreaFactor = 4000
			Temperature=@Temp@  
			Mobility( 
			DopingDep 
			HighFieldSat 
			Enormal 
			)
			Recombination( 
			SRH (TempDependence DopingDependence) 
			Avalanche(UniBo) 
			Auger
			)
			
			EffectiveIntrinsicDensity( OldSlotboom )
			#if "@Fluence@" != "0"    	
			Traps ( 
			(Donor  Conc=@ConcDon1@ Level EnergyMid=0.23 fromCondBand eXsection=2.3e-14 hXsection=2.3e-15 Add2TotalDoping)
			(Acceptor Conc=@ConcAcc1@ Level EnergyMid=0.42 fromCondBand eXsection=1.0e-15 hXsection=1.0e-14 Add2TotalDoping)
			(Acceptor Conc=@ConcAcc2@ Level EnergyMid=0.46 fromCondBand eXsection=7.00e-14 hXsection=7.00e-13 Add2TotalDoping)
			)
			#endif
		}
		
		Physics(MaterialInterface="Silicon/Oxide"){
			Recombination(surfaceSRH)
			Traps(
			(FixedCharge Conc=@Qox@)
			(Acceptor Conc=@Dit_acc@ Uniform EnergyMid=0.83 EnergySig=0.58	 fromValBand   eXsection=1e-16 hXsection=1e-15 Add2TotalDoping)    
			(Donor Conc=@Dit_don@ Uniform EnergyMid=0.27 EnergySig=0.54 fromValBand eXsection=1e-15 hXsection=1e-16 Add2TotalDoping)
			)
		}		
		
	\end{lstlisting}

	\section{Declaration}
	Patrick Asenov and Panagiotis Assiouras would like to declare that they have contributed equally to this work.

	\section{Acknowledgments}
	
	The authors would like to thank the Tracker Group of the CMS Collaboration for providing the MOS capacitor test structures.

\begin{wrapfigure}{l}{3.0cm}
\vspace{-10pt}
\includegraphics[width=3.0cm]{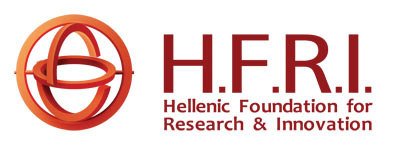}
\end{wrapfigure} 
This work is supported by the  Grant No 5029538 from the Structural Funds, European Regional Development Funds (ERDF) and European Structural Funds (ESF), Greece and 
by the Hellenic  Foundation for Research and Innovation (HFRI) Greece, under the  HFRI PhD Fellowship grant (Fellowship Number: 26)

%------------------------------------------


\begin{thebibliography}{99}
\bibitem{a}
G. Apollinari et al.,
\emph{High-Luminosity Large Hadron Collider (HL-LHC): Technical Design Report V. 0.1},
\emph{CERN Yellow Rep. Monogr.} {\bf 4} (2017) 1-516
[{\tt CERN-2017-007-M}].

%\bibitem{b}
%\emph{Corporate Profile},
%\emph{Hamamatsu Photonics}, \href{https://www.hamamatsu.com/eu/en/our-company/corporate-profile/index.html} {https://www.hamamatsu.com/eu/en/our-company/corporate-profile/index.html} .

\bibitem{b}
H. Simon,
\emph{Chapter 1: The Mystique of the Hidden Champions. Hidden Champions of the Twenty-First Century: The Success Strategies of Unknown World Market Leaders},
\emph{London: Springer Science+Business Media}, p.4 ISBN 978-0-387-98147-5.

\bibitem{picker}
K. Tsien, R. Robbins,
\emph{A comparison of a cobalt-60 teletherapy unit and a 2-MEV Van de Graff x-ray generator on the basis of physical measurements},
\emph{Radiology} 1958 Apr; {\bf{70(4)}}: 486-500; discussion 501-2.

%\bibitem{c}
%\emph{FC65-G/FC65-P - Standard reference chambers},
%\href{https://www.iba-dosimetry.com/product/fc65-g-fc65-p-ionization-chambers/} {https://www.iba-dosimetry.com/product/fc65-g-fc65-p-ionization-chambers/} .

\bibitem{c}
L. Burigo, S. Greilich,
\emph{Impact of new ICRU 90 key data on stopping-power ratios and beam quality correction factors for carbon ion beams},
\emph{Phys Med Biol.}, 2019 Sep 23; {\bf{64(19)}} :195005.

%\bibitem{d}
%\emph{Bureau International des Poids et Mesures},
%\href{https://www.bipm.org/en/about-us/} {https://www.bipm.org/en/about-us/} .

\bibitem{d}
C.H. Page, P. Vigoureux, eds.,
\emph{The International Bureau of Weights and Measures 1875-1975: NBS Special Publication}, {\bf{420}}, Washington, D.C.: \emph{National Bureau of Standards} pp. 26-27.

\bibitem{e}
\emph{Total dose steady-state irradiation test method},
\emph{ESCC Basic Specification No.} {\bf 22900} .

\bibitem{mgs}
IMS - Innovation \& Measurement Systems,
\emph{Micro-Sized gamma Spectrometer, MGS Series} datasheet.

\bibitem{nist}
J. H. Hubbell and S. M. Seltzer,
\emph{X-Ray Mass Attenuation Coefficients}, \emph{NIST Standard Reference Database 126}, {\bf NISTIR 5632}, DOI: \href{https://dx.doi.org/10.18434/T4D01F} {https://dx.doi.org/10.18434/T4D01F}

\bibitem{us_doh}
Bureau of Radiological Health and the Training Institute, Environmental Control Administration,
\emph{Radiological Health Handbook}, {\bf January 1970,page 139}, DOI: \href{https://doi.org/10.2172/4708654} {https://doi.org/10.2172/4708654}

\bibitem{root}
R. Brun and F. Rademakers,
\emph{ROOT - An Object Oriented Data Analysis Framework}, 
Proceedings AIHENP'96 Workshop, Lausanne, Sep. 1996, Nucl. Instr. \& Meth. in Phys. Res. {\bf A 389} (1997) 81-86.

\bibitem{f}
Bart V. Van Zeghbroeck,
\emph{Principles Of Semiconductor Devices And Heterojunctions},
Prentice Hall, December 1st 2009.

\bibitem{mir}
G. S. Baranenkov,
\emph{Problems in mathematical analysis},
Moscow: Mir Publishers, 1976.

\bibitem{mosc}
F. Moscatelli et al., 
\emph{Effects of Interface Donor Trap States on Isolation Properties of Detectors Operating at High-Luminosity LHC},
\emph{IEEE Trans. Nucl. Sci. 64} {\bf{(8)}} (2017) 2259–2267.

\bibitem{mor}
A. Morozzi et al.,
\emph{TCAD advanced radiation damage modeling in silicon detectors}
\emph{Volume {\bf{373}} - The 28th International Workshop on Vertex Detectors (VERTEX 2019)}

\bibitem{lindstrom}
Gunnar Lindstr$\mathrm{\ddot{o}}$m,
\emph{Radiation Damage in Silicon Detectors}, 
invited talk presented at the 9th European Symposium on Semiconductor Detectors, Schloss Elmau, Germany, June 23-27, 2002; accepted for publication in Nucl. Instr. and Meth. A.

\bibitem{fretwurst}
E. Fretwurst et al.,
\emph{Bulk damage effects in standard and oxygen-enriched silicon detectors induced by 60Co-gamma radiation},
\emph{Nuclear Instruments and Methods in Physics Research Section A: Accelerators, Spectrometers, Detectors and Associated Equipment} {\bf Volume 514}, Issues 1-3, 21 November 2003, Pages 1-8.

\bibitem{h}
A. Chilingarov,
\emph{Generation current temperature scaling},
\emph{RD50 and EP Technical Note}
[{\tt RD50 TN 2011-01}].

\bibitem{TCAD}
\emph{\url{https://www.synopsys.com/silicon/tcad.html}"},year =2019, [Online; accessed 01-July-2019]

\bibitem{Moll}
\emph{M. Moll}. 
\emph{Radiation Damage in Silicon Particle Detectors Microscopic Defects and Macroscopic Properties}. PhD Thesis, University of Hamburg, 1999. DESY-THESIS-1999-040.

\bibitem{Schwandt},
\emph{J. Schwandt and E. Fretwurst and E. Garutti and R. Klanner and C. Scharf and G. Steinbrueck},
\emph{A new model for the TCAD simulation of the silicon damage by high fluence proton irradiation}, 2018 IEEE Nuclear Science Symposium and Medical Imaging Conference Proceedings (NSS/MIC), 2018.

\bibitem{Passeri1}
D. Passeri, P. Ciampolini, G. M. Bilei and F. Moscatelli, 
\emph{Comprehensive modeling of bulk-damage effects in silicon radiation detectors}
\emph{IEEE Transactions on Nuclear Science}, vol. 48, no. 5, pp. 1688-1693, Oct. 2001, doi: 10.1109/23.960358.


\bibitem{Petasecca}
M. Petasecca, F. Moscatelli, D. Passeri and G. U. Pignatel, 
\emph{Numerical Simulation of Radiation Damage Effects in p-Type and n-Type FZ Silicon Detectors} 
\emph{IEEE Transactions on Nuclear Science}, vol. 53, no. 5, pp. 2971-2976, Oct. 2006, doi: 10.1109/TNS.2006.881910.

\bibitem{Dalal}
\emph{Dalal, R and Bhardwaj, A and Ranjan, K and Moll, M and Elliott-Peisert, A"}
\emph{Combined effect of bulk and surface damage on strip insulation properties of proton irradiated n$^{+}$-p silicon strip sensors}, JINST vol. 9, P04007, 2014.


\bibitem{mosc}
F. Moscatelli et al., 
\emph{Effects of Interface Donor Trap States on Isolation Properties of Detectors Operating at High-Luminosity LHC},
\emph{IEEE Trans. Nucl. Sci. 64} {\bf{(8)}} (2017) 2259–2267.

\bibitem{mor}
A. Morozzi et al.,
\emph{TCAD advanced radiation damage modeling in silicon detectors}
\emph{Volume {\bf{373}} - The 28th International Workshop on Vertex Detectors (VERTEX 2019), PoS (Vertex2019) 050, https://doi.org/10.22323/1.373.0050}

\bibitem{mor1}
A. Morozzi et al., 
\emph{Characterization of irradiated p=type silicon detectors for TCAD surface radiation damage model validation},
\emph{2020, JINST, 15, C01029, https://doi.org/10.1088/1748-0221/15/01/C01029 }.

%\bibitem{MOSCATELLI2019198}
%F. Moscatelli et al,
%\emph{\say{Analysis of surface radiation damage effects at HL-LHC fluences: Comparison of different technology options}},
%\emph{\say{Nuclear Instruments and Methods in Physics Research Section A: Accelerators, Spectrometers, Detectors and Associated Equipment}},
%\emph{volume = \say{\bf{924}}}, pages \say{198 - 202}, year = \say{2019}.

%\bibitem{Moscatelli3}
%F. Moscatelli et al,
%\emph{\say{Combined Bulk and Surface Radiation Damage Effects at Very High Fluences in Silicon Detectors: Measurements and TCAD Simulations}}
%\emph{IEEE Transactions on Nuclear Science}
%vol. 63, no. 5, pp. 2716-2723, Oct. 2016, doi: 10.1109/TNS.2016.2599560.

\bibitem{Nicollian}
E.H. Nicollian and J. R. Brews, 
\emph{MOS (Metal Oxide Semiconductor), Physics and technology}, 
\emph{John Wiley and Sons}, 1982, pp. 319-356.

%\bibitem{Passeri2}
%\emph{D. Passeri and F. Moscatelli and A. Morozzi and G.M. Bilei},
%\emph{Modeling of radiation damage effects in silicon detectors at high fluences HL-LHC with Sentaurus TCAD},
%\emph{\say{Nuclear Instruments and Methods in Physics Research Section A: Accelerators, Spectrometers, Detectors and Associated Equipment}},vol. \say{824}, pp 443 - 445, year = \say{2016},issn %= \say{0168-9002}.


\bibitem{Sentaurus}
\emph{Sentaurus Device User Guide}, Synopsys, Imnc.


\end{thebibliography}
\end{document}